\documentclass[review]{elsarticle}
\usepackage{graphicx}
\usepackage{bm}
\usepackage[utf8]{inputenc}
\usepackage[T1]{fontenc}
\usepackage{mathptmx}
\usepackage{bbold}
\usepackage{color}
\usepackage{amssymb}
\journal{Journal of Physics A: Mathematical and Theoretical}
\begin{document}
\begin{frontmatter}
\title{Exact closed forms for the transmittance of
    electromagnetic waves in one-dimensional anisotropic
    periodic media} 

\author[buap]{J. C. Torres-Guzmán \corref{cor1}}%
\ead{jtorres@ifuap.buap.mx}
\author[buap]{A. Díaz de Anda}
\author[buap]{J. Arriaga}
\cortext[cor1]{Corresponding author}
\address[buap]{Instituto de Física, Benemérita Universidad Autónoma de
  Puebla, P.O. Box J-48 72570, Puebla, Puebla, Mexico}

\begin{abstract}
  In this work, we obtain closed expressions for the transfer matrix
  and the transmittance of electromagnetic waves propagating in finite
  1D anisotropic periodic stratified media with an arbitrary number of
  cells. By invoking the Cayley-Hamilton theorem on the transfer
  matrix for the electromagnetic field in a periodic stratified media
  formed by $N$ cells, we obtain a fourth-degree recursive relation
  for the matrix coefficients that defines the so-called Tetranacci
  Polynomials. In the symmetric case, corresponding to a unit-cell
  transfer matrix with a characteristic polynomial where the 
  coefficients of the linear and cubic terms are equal, closed
  expressions for the solutions to the recursive relation, known as
  symmetric Tetranacci Polynomials, have recently been derived,
  allowing us to write the transfer matrix and transmittance in a
  closed form. We show as sufficient conditions that the
  $4\times4$ differential propagation matrix of each layer in the
  binary unit cell, $\mathbb\Delta$, $a)$ has eigenvalues of the form
  $\pm~p_1$, $\pm~p_2$, with $p_1\ne p_2$, and $b)$ its off-diagonal
  $2\times2$  block matrices possess the same symmetric structure in
  both layers. Otherwise, the recursive relations are still solvable
  for any $4\times4$-matrix and provide an algorithm to compute the
  $N$-th power of the transfer matrix without carrying out explicitly
  the matrix multiplication of $N$ matrices. We obtain analytical
  expressions for the dispersion relation and transmittance, in closed
  form, for two finite periodic systems: the first one consists of two
  birefringent uniaxial media with their optical axis perpendicular to
  the $z$-axis, and the second consists of two isotropic media subject
  to an external magnetic field oriented along the $z$-axis and
  exhibiting the Faraday effect. Our formalism applies also to lossy
  media, magnetic anisotropy or optical activity.  

\end{abstract}
\end{frontmatter}

\section{Introduction}

Unlike travelling or standing waves, wave propagation in infinite
periodic media allows frequencies that fall into continuous
bands, separated by forbidden gaps. This fact, first noted by Kronig
and Penney, is the foundation for the modern theory of solids
\cite{Kronig-Penney}. The band structure is at the core of solid-state
physics and has been studied in many different systems, quantum and
classical. The systems range from the nanometric scale in photonics to
oceanographic waves, covering a range in size of more than a dozen
orders of magnitude.   

In the study of infinite periodic systems, the problem is dramatically
simplified thanks to Bloch's theorem \cite{Bloch}. However, in 
practice only a few systems can be considered infinite
periodic. Therefore, finite periodic systems are more difficult to
analize, since Bloch's theorem does not apply. Fortunately, the finite
periodic case, involving $2\times2$-matrices, can be solved
analytically for an arbitrary number of unit cells, $N$, first
demonstrated in the optical case by Abèles \cite{Abeles} in 1950 and
extended to the quantum context \cite{Cvetic} in 1981 and later by
others \cite{Erdos,Kiang}. Thus, for one-dimensional waves in locally 
periodic (finite periodic) media 
\cite{Griffiths2001,YarivBook,Yeh-1977,Bandelow-1993,Lekner-1994,Bendikson-1996,
Vadim-1999}, analytical expressions for quantities
like the transmittance are available in terms of the Chevychev
polynomials of the second kind \cite{Abromowitz1965}.      

Since then, light propagation in modern optical systems has become a
central issue. Thin films, periodic layered media, and other
stratified systems exhibit novel and attractive features in
technological applications as anti-reflection coatings
\cite{Hiller-2002,Xi-2007}, x-ray mirrors \cite{Boher-1992,Spiller-1994},
transistors \cite{Nomura-2003,Street-2009}, thin-film photovoltaics
\cite{Chopra-2004,Shin-2013}, photonic crystals
\cite{Joannopoulos-1995}, among others.
In particular, layered optical periodic media finds
  application in the development of photonic topological insulators 
\cite{Hasan-2010,Kane-2005,Joannopoulos-2014,Khanikaev-2013,Schomerus-2013,Rechtsman-2013,Weimann-2017},
as they possess robust edge states, which persist in the presence of
perturbations and make them ideal for the transfer or storage of
energy or information. Even realizations of robust edge states
in one-dimensional structures
\cite{Fefferman-2014,Thorp-2016,Levy-2017,Xiao-2014,Choi-2016}, have
been reported recently.  

Many of the above active and passive optical devices require the
epitaxial growth of anisotropic thin films. Therefore, the design of
these devices relies on powerful mathematical and numerical techniques
to adequately describe light propagation in anisotropic layered media.   
In this direction and from a theoretical point of view, the Abèles
formalism was extended to a $4\times4$-matrix technique \cite{Teitler}
to describe the propagation and reflection by stratified anisotropic
dielectric media, and later extended to include magnetic anisotropy
and optical activity \cite{Berreman-1972}. In this formalism, the
transfer matrix of an anisotropic layer is expressed in terms of its 
eigenvalues and eigenvectors. However, the solution of the associated
eigenvalue problem is left out. This eigenvalue problem is then
solved and incorporated into a new formalism to
describe the electromagnetic propagation in non-magnetic arbitrarily
birefringent layered media \cite{Yeh-1979,Yeh-1980}. By taking
advantage of Bloch's theorem and by carrying out the matrix
multiplication of the whole layered system, this new formalim allowed
the study of the band structure and Bloch waves in a periodic
birefringent media. These works have led to the
  development of several numerical methods to describe light
propagation in anisotropic stratified systems
\cite{Lin-1984,Li-1988,Nikolai}. Recently, a 
different approach has been derived \cite{Stepan-2022} to study
isotropic and anisotropic alternating layers with one or two
anisotropic defect layers in the framework of the temporal
coupled-mode theory. Despite these commendable efforts,
closed-form analytical expressions, e.g. for
  transmittance, are not yet available.

Among the pioneering efforts on the formulation of
the transfer matrix in the context of finite elastic periodic
systems, we should mention the work of Lin and McDaniel
\cite{Lin-1969}, who used the Cayley-Hamilton (C-H) theorem
\cite{Griffiths2001,LangBook,BarutO-1994,BarutSU-1994}, to obtain an 
expression for the $4\times4$-transfer matrix of $N$ cells in the
propagation of bending oscillations. Unfortunately, this expression
is invalid for degenerate eigenvalues of the unit-cell matrix
\cite{Laufer-1997}. 
We have recently shown \cite{Torres-2024} by a similar procedure as
that of Ref.~\cite{Lin-1969}, that the $N$-th power of a
$4\times4$-matrix can be reduced to the calculation of a set of four 
functions, recursively defined, instead of multiplying the matrices,
as was the original goal in Ref.~\cite{Lin-1969}. Such recursive
relations define the Tetranacci Polynomials (TP) \cite{Soykan2020} and
are solvable for an arbitrary number of cells, $N$
\cite{Torres-2024}. Furthermore, we have shown that
in free oscillations of Timoshenko-Ehrenfest beams
\cite{Torres-2024}, the eigenvalues of the transfer matrix
corresponding to each layer composing the unit cell are
non-degenerate. Hence, we obtained the recursive relationship for
symmetric TP, which allows us to use the recently derived 
closed-form expressions obtained in the context of a
finite Kitaev chain \cite{Leumer2020,Leumer2021,Leumer-2023} to
write the transfer matrix of $N$ cells in a closed-form.
Moreover, these expressions account for the degeneracy of the
eigenvalues of the unit-cell transfer matrix \cite{Laufer-1997}. In
this work, we obtain the sufficient conditions the optical
$4\times4$-differential propagation matrix of each layer in the unit
cell must fulfil to obtain closed forms, in terms of symmetric TP's,
for the transmittance in finite 1D anisotropic periodic stratified
media.   

The structure of the paper is as follows: In Section 2, we review 
the quadrivector formulation to describe the propagation of waves in 
optical stratified media. In Section 3, we obtain a set of four
recursive relations to calculate the $N$-th power of the transfer
matrix by using the C-H theorem. In Section 4, we discuss the
restrictions on the $4\times4$ differential propagation matrix to 
fulfil the symmetry condition that allows the use of symmetric
TP's. In Section 5, we obtain the dispersion relation
without using explicitly Bloch's theorem. In Section
6, we obtain closed forms for the $N$-th power of the transfer matrix
and the transmittance of the finite periodic system. In Section 7, we
give our conclusions.

\section{$4\times4$-Matrix formulation}
We are interested in a locally periodic anisotropic media consisting
of $N$ cells. First, we review the electro-magnetic field in each
layer. The specific nature of each layer can be different in the more
general case. In special cases, exact results can be obtained. 
Thus, a constitutive relation between the vectors $\mathbf
C=(D_x,D_y,D_z,B_x,B_y,B_z)$ and $\mathbf
G=(E_x,E_y,E_z,H_x,H_y,H_z)$, where $E_j$, $D_j$, $B_j$ and $H_j$,
$j=x,y,z,$ are the cartesian components of the electromagnetic-field
vectors, can be generally expressed  as  
\begin{equation}
  \mathbf C=\mathbf M \mathbf G,
\end{equation}
where the matrix $\mathbf M$, is the so-called optical matrix and
carries all the information about the anisotropic optical properties
of the medium \cite{Berreman-1972}. The optical matrix can be
partitioned as
\begin{equation}
  \mathbf M=\left(\begin{array}{cc}
    \mathbf \epsilon&\mathbf \rho\\
    \mathbf \rho'&\mathbf \mu \end{array}\right), 
\end{equation}
where as usual, $\mathbf \epsilon$ and $\mathbf \mu$ are the
permitivity and permeability tensors, respectively, and $\mathbf \rho$
and $\mathbf\rho'$ are the optical rotation tensors
\cite{Berreman-1972}. Consider the propagation of a monochromatic
plane wave of frequency $\omega$ obliquely incident
int the $xz$-plane from an isotropic medium ($z<0$) onto a periodic
anisotropic layered media of $N$ cells ($z>0$) stratified along the
$z$-axis, in which $\mathbf M$ is a function only of $z$. In this
case, the $x$-component of the propagation vector, $k_x$, is constant
and there is no $y$ component in any layer of the structure, 
\begin{equation}
  k_x=\frac{\omega}{c}n_0\sin \theta_0,
\end{equation}
where $c$, $n_0$, and $\theta_0$ are the free-space wave velocity,
the refractive index of the isotropic ambient and the angle of
incidence, respectively.   
Then, the Maxwell's equations in an anisotropic layer can be cast into
a $4\times4$ matrix form \cite{Berreman-1972}
\begin{equation}\label{difPsi}
\frac{\partial \bm \Psi}{\partial z}=i\omega \mathbb \Delta \bm \Psi,
\end{equation}
where 
\begin{equation}\label{Psi}
  \bm \Psi=\left(\begin{array}{c}
  E_x\\H_y\\E_y\\-H_x\end{array}\right),
\end{equation}
is the generalized quadrivector field, and the $4\times4$-matrix
$\mathbb\Delta$ is function of the elements of the $6\times6$-optical
matrix $\mathbf M$. The details to construct $\mathbb~\Delta$ from
$\mathbf M$ can be found in Refs.~\cite{Berreman-1972,Azzam}.   
In the general case of a stratified anisotropic structure, $\mathbf M$ 
is some arbitrary function of $z$ and the Eq.~(\ref{difPsi}) does not,
in general, have an analytical solution. In the special case when
$\mathbf M$ is independent of $z$, Eq.~(\ref{difPsi}) is
directly integrable to yield
\begin{equation}
 \bm \Psi(z+\ell)=\mathbf T(\ell)\bm\Psi(z),
\end{equation}
where $\mathbf T(\ell)=e^{i\omega\ell\mathbb \Delta}$, which can be expanded
in a power series and the sum carried out analytically in a few cases
only. Analytical expressions for $\mathbf T(\ell)$ can be
determined by analyzing the four plane-wave solutions of
Eq.~(\ref{difPsi}). We address this case in Section 4. Here, we pause
the $4\times4$-matrix formulation to introduce the C-H theorem, upon
which our formalism relies to obtain the transfer matrix of an
anisotropic periodic media.    

\section{The Cayley-Hamilton theorem and the Tetranacci polynomials
  recurrence relation}\label{cayley}

Consider a locally periodic anisotropic media consisting of $N$
cells. Each unitary cell consists of two anisotropic layers of
thickness $\ell_1$ and $\ell_2$, respectively.
Let us choose the unitary cell such that the frontiers are at
$z=z_0$ and $z=z_2$, and the interface between the two layers is at
$z=z_1$, where $z_0<z_1<z_2$. Then the EM-quadrivector, $\bm\Psi$, at
$z=z_1$ can be written in terms of its value at $z=z_0$,
\begin{equation}
\bm  \Psi(z_1)=\mathbf T_1(\ell_1)\bm\Psi(z_0),
\end{equation}
where $\mathbf T_1(\ell_1)$ is the transfer matrix for the first layer.
Similarly,
\begin{equation}
\bm  \Psi(z_2)=\mathbf T_2(\ell_2)\bm\Psi(z_1).
\end{equation}
At $z=z_1$, we demand continuity in the parallel components of the EM
field, i.e. continuity on $\bm\Psi$. Thus, the transfer matrix for the
unitary cell, $\mathbf T_c$, is given by  
\begin{equation}\label{eqforT}
  \mathbf T_c =\mathbf T_2(\ell_2)\mathbf T_1(\ell_1).
\end{equation}
The transfer matrix for $N$ cells is then given by $\mathbf{T}_c^N$.  
Now we invoke the C-H theorem
\cite{Griffiths2001,BarutO-1994,BarutSU-1994} to
calculate $\mathbf{T}_c^N$. It states that any matrix, $\mathbf T_c$,  
obeys its own characteristic polynomial equation
\begin{equation}\label{charm}
  \mathbf T_c^4-\eta \mathbf T_c^3+\zeta \mathbf T_c^2 -\tau \mathbf
  T_c +\delta_0\mathbf I=\mathbb{0},
\end{equation}
where
\begin{eqnarray}
  \eta&=&\mbox{Tr}(\mathbf T_c),\label{eta}\\
  \zeta&=&\frac{1}{2}\left(\eta^2-\mbox{Tr}(\mathbf
  T_c^2)\right),\label{zeta}\\ 
  \tau&=&\frac{1}{6}\left(\eta^3-3\eta\mbox{Tr}(\mathbf T_c^2)+
  2\mbox{Tr}(\mathbf T_c^3)\right),\label{Delta}\\
  \delta_0&=&\mbox{det}(\mathbf T_c),\label{delta0}
\end{eqnarray}
where Tr and det denote the trace and the determinant of a matrix,
respectively. 
$\mathbf I$ and $\mathbb{0}$ are the identity and null
$4\times4$-matrices, respectively. 
As in Refs.~\cite{Lin-1969,Torres-2024}, this
equation can be then used to calculate any power, $N$, of 
$\mathbf T_c$ 
\begin{equation}\label{powerm}
  \mathbf T_c^N=\alpha(N-1) \mathbf T_c^3-\beta(N-2)\mathbf T_c^2
  +\gamma(N-3)\mathbf T_c -\delta(N-4)\mathbf I, \quad \mbox{for }
  N\ge5. 
\end{equation}
where the functions $\alpha(n)$, $\beta(n),$ $\gamma(n)$, and
$\delta(n)$, are recursively defined by \cite{Torres-2024},
\begin{eqnarray}
  \alpha(n)&=&\eta\alpha(n-1)-\beta(n-2),\label{alpha}\\
  \beta(n-1)&=&\zeta\alpha(n-1)-\gamma(n-3),\label{beta}\\
  \gamma(n-2)&=&\tau\alpha(n-1)-\delta(n-4),\label{gamma}\\
  \delta(n-3)&=&\delta_0\alpha(n-1)\label{delta}.
\end{eqnarray}
Equation (\ref{powerm}) can be demonstrated by mathematical
induction. Eqs.~(\ref{alpha})-(\ref{delta}) are valid for $n\ge4$ with   
initial values $\alpha(3)=\eta$, $\beta(2)=\zeta$, $\gamma(1)=\tau$
and $\delta(0)=\delta_0$, corresponding to the case $N=4$. The
powers, $N=2,3$, have to be done by matrix multiplication, but this is  
computationally cheap. 
Substitution of Eqs.~(\ref{beta})-(\ref{delta}) into Eq.~(\ref{alpha})
yields an Eq. for $\alpha(n)$ solely \cite{Torres-2024}
\begin{equation}\label{recursion}
\alpha(n)=\eta\alpha(n-1)-\zeta\alpha(n-2)+\tau\alpha(n-3)
-\delta_0\alpha(n-4),    
\end{equation}
valid for $n\ge7$. Such recurrence relationship defines the Tetranacci  
polynomials \cite{Soykan2020}, with coefficients $\eta, \zeta,\tau$,
and $\delta_0$. With the initial values, $\alpha(3)$,
$\beta(2)$, $\gamma(1)$ and $\delta(0)$, and
Eqs.~(\ref{alpha})-(\ref{delta}), we can obtain the other seminal
values, $\alpha(j)$, $j=4,5,6,$ to iterate Eq.~(\ref{recursion}) and
obtain $\alpha(N-1)$. 
Then, the values $\alpha(n)$, with $n=N-1, N-2,N-3,N-4$ can be
substituted back into Eqs.~(\ref{alpha})-(\ref{delta}) once again, to
obtain the coefficients $\alpha(N-1)$, $\beta(N-2)$,
$\gamma(N-3)$, and $\delta(N-4)$, required in Eq.~(\ref{powerm}) to
finally obtain $\mathbf T_c^N$. Therefore, the computational cost
is significantly reduced since no matrix multiplication is involved  
beyond the one implicit in $\mathbf T_c^2$ and $\mathbf T_c^3$.
Also, the recurrence relationship of Eq.~(\ref{recursion}) is valid
even for degenerate eigenvalues of the transfer matrix $\mathbf T_c$
\cite{Laufer-1997}. This iterative algorithm is one of the
methods implemented in this work and we will refer to it as Method A
\cite{Torres-2024}.  

\section{The traceless matrix condition} 

Further simplifications to calculate the $N$-th power of the transfer
matrix $\mathbf T$ can be achieved if the matrix, $\mathbb \Delta$,
in addition to being constant and independent of $z$, has a
characteristic polynomial that is an even function
\cite{BarutO-1994,BarutSU-1994,Torres-2024}. In other words, if $p$ is
an eigenvalue of $\mathbb \Delta$, then $-p$ is also an
eigenvalue. Hence, its eingenvalues are $(p_1,p_2,-p_1,-p_2)$, 
with $p_1\ne~p_2$. Although this condition seems arbitrary and very
restrictive, it is not. This condition is consistent with 
having identical waves propagating in both directions $\pm z$, a
situation encountered in many physical systems beyond the particular
cases addressed here. Consider the ansatz for the solution of
Eq.~(\ref{difPsi}), 
\begin{equation}\label{solPsi}
\bm\Psi(z)=\bm\Psi_\ell(0)e^{ik_\ell z}, \quad \ell=1,2,3,4,
\end{equation}
that accounts for four distinct plane-wave solutions.
The substitution of Eq.~(\ref{solPsi}) into Eq.~(\ref{difPsi}) yields
the matrix eigenvalue equation, 
\begin{equation}\label{eigenvalor}
(k_z\mathbf I-\omega\mathbb \Delta)\bm \Psi(0)=0,
\end{equation}
whose eigenvalues, $k_\ell$ (we have dropped the
  subindex $z$ to simplify nomenclature), are proportional to the
roots, $p_\ell$, of the characteristic polinomial of $\mathbb\Delta$,
with $\omega$ the proportionality constant, since  $k_\ell$ may be
analytically obtained by solving the quartic polynomial equation, 
\begin{equation}\label{caract}
  \mbox{det}[k_z\mathbf{I}-\omega\mathbb \Delta]=0.
\end{equation}
Therefore $k_3=-k_1$ and $k_4=-k_2$, corresponding to two waves
propagating in both directions $\pm z$. 

According to the discussion above, the even parity of the
characteristic polynomial implies that both its linear and cubic   
coefficients, $\eta_{\mathbb\Delta}$ and $\tau_{\mathbb\Delta}$,
respectively, vanish,
i.e. $\eta_{\mathbb\Delta}=\tau_{\mathbb\Delta}=0$. In terms of
traces, according to Eqs.~(\ref{eta})-(\ref{Delta}) 
applied to $\mathbb\Delta$, we have
Tr~$\mathbb\Delta=\mbox{ Tr}~\mathbb\Delta^3=0$,   
i.e. both $\mathbb\Delta$ and $\mathbb\Delta^3$ matrices are traceless 
\cite{BarutO-1994,BarutSU-1994,Torres-2024}. The
corresponding non-degeneracy condition, $p_1\ne~p_2$, has two
implications: the first one is that the independent term,
$\mbox{det }\mathbb\Delta$,  must be different from zero,
i.e. $|\mathbb\Delta|\ne0$, to avoid the trivial and degenerate
solution, $p=0$. The second implication is that according to
Eq.~(\ref{zeta}) applied to $\mathbb\Delta$,
$(\mbox{Tr~}\mathbb\Delta^2)^2\ne16\mbox{ det~}\mathbb\Delta$.

We show in Appendix A, that two types of $\mathbb\Delta$-matrices can  
be distinguished for optical systems that fulfill the above
conditions according to the symmetry of its off-diagonal $2\times2$
block matrices. If the two constituents of the unit cell have the same 
symmetric structure, then we have $\eta=\tau$, i. e. the
characteristic polynomial, Eq.~(\ref{charm}), is symmetric. We
demonstrate this in Appendix B, by noticing from 
the reciprocal polynomial that if $\eta=\tau$, then $\mbox{Tr~}\mathbf
T_c=\mbox{Tr~}\mathbf T_c^{-1}$ \cite{Torres-2024}.  
Under the category of type I, we can identify the case of the uniaxial
orthorhombic crystal with its optical axis perpendicular to the axis 
of propagation and rotated from the plane of incidence
\cite{Azzam,Sosnowski}, or the Faraday rotator of light subjected to a  
magnetic field in the $z$ direction
\cite{Berreman-1972,Born}. Moreover, under the category of type II, we 
can find the Drude model for an isotropic optically active medium
\cite{Berreman-1972,Drude-1965}, or other similar models for the same
 phenomenon \cite{Treyakov-1998,Oldano-1999,Wade-2020}. Unfortunately,  
as we show in Appendix B, if the constituents of the unit cell have 
different symmetric structure, then $\eta\ne\tau$. Even so, we can
have closed expressions for the transmittance of periodic optical
systems consisting of uniaxial crystals, Faraday rotators or
combinations of these. The same applies for periodic optical systems
consisting of layers with natural optical activity.    

The general solution for $\bm\Psi$ is a linear combination of the
eigenvectors, $\vec V_j$, $j=1,2,3,4$, contained as columns in the
matrix $\mathbb V=(\vec V_1,\vec V_2,\vec V_3,\vec V_4)$,  
\begin{equation}\label{solforF}
\bm \Psi(z)=\mathbb{V} \mathbb\lambda(z)\vec{A}, 
\end{equation}
where 
\begin{equation}\label{eigenvalues}
  \mathbb \lambda(z)=\mbox{diag}(e^{ik_1z},e^{ik_2z},e^{-ik_1z},e^{-ik_2z}),
\end{equation}
i.e., $\mathbb\lambda(z)$ is a diagonal matrix,
and $\vec A=(A_1,A_2,A_3,A_4)^T$, where $A_j$, $j=1,...,4$, are
arbitrary constants. If $\bm\Psi$ is known at $z=z_0$ in a layer, then 
the value of $\bm\Psi$ at any $z$ value in the same layer, can be
obtained by solving for $\vec A$ from
$\bm\Psi(z_0)=\mathbb{V}\mathbb\lambda(z_0)\vec{A}$ and then
substituting into Eq.~(\ref{solforF}) to obtain 
\begin{equation}\label{tmforF}
\bm\Psi(z)=\mathbb{V}\mathbb\lambda(z-z_0)\mathbb{V}^{-1}\bm\Psi(z_0),   
\end{equation}
which provides the transfer matrix, $\mathbf T$, for the quadrivector
$\bm\Psi$ in the layer in question \cite{Berreman-1972},
\begin{equation}\label{tm}
\mathbf T(z-z_0)=\mathbb{V}\mathbb\lambda(z-z_0)\mathbb{V}^{-1}.
\end{equation}
From Eqs.~(\ref{eigenvalues}) and (\ref{tm}), we also obtain
$\mbox{det}(\mathbf T)=1$ \cite{Hall-2015}.
We emphasize that Eq.~(\ref{tm}) is valid only if $k_1\ne~k_2$,
otherwise $\mathbb\Delta$ might not have four independent
eigenvectors. It is precisely for isotropic media when the case,
$k_1=~k_2$, might occur, and therefore we exclude
this possibility. However, from a practical point of view, we can
recover the case of an isotropic layer to any degree of approximation
by taking the anisotropy arbitrarily small, such that the above
expression is still valid to avoid discontinuous solutions 
\cite{Nikolai,Xu-2000,Zhang-2015}.  

Notice also, the transfer matrix, $\mathbf T$, is the exponential of
the $\mathbb\Delta$ matrix \cite{Hall-2015}
\begin{equation}\label{tmexp}
\mathbf T(z-z_0)=e^{i\omega\mathbb\Delta(z-z_0)},
\end{equation}
and therefore, the transfer  matrix of the unit cell, $\mathbf T_c$,
is then given by   
\begin{equation}\label{tmexpcell}
\mathbf T_c(\ell)=e^{i\omega\mathbb\Delta_2\ell_2}e^{i\omega\mathbb\Delta_1\ell_1},  
\end{equation}
which is equal to the exponential of the sum of the matrix exponents
only if $\mathbb\Delta_1$ and $\mathbb\Delta_2$ commute. However, this
condition is generally not satisfied and then the transfer matrix,
$\mathbf T_c$, can be computed by means of the Lie product formula
\cite{Hall-2015} and taking a large number of terms to approximate
it. Alternatively, if $\mathbb\Delta_1$ and $\mathbb\Delta_2$ are
sufficiently small matrices, then $\mathbf T_c=e^{i\omega \mathbf Q}$,
where $\mathbf Q$ may be computed as a series in commutators of
$\ell_1\mathbb\Delta_1$ and $\ell_2\mathbb\Delta_2$ by using the
Baker-Campbell-Hausdorff formula \cite{Highman}. This approach is used
in Ref.~\cite{Broer-2023} to investigate the effects of chirality in
the Casimir torque.       

\section{The dispersion relation}

When the symmetric condition $\eta=\tau$ is substituted into
Eq.~(\ref{recursion}), the recursion relation for the symmetric
Tetranacci polynomials is obtained,
\begin{equation}\label{symrecursion}
\alpha(n)=\eta\alpha(n-1)-\zeta\alpha(n-2)+\eta\alpha(n-3)-\alpha(n-4),
\quad n\in \mathbb Z, 
\end{equation}
in terms of its initial values $\alpha(j)$, $j=-2,-1,0,1$, and
complex coefficients, $\eta$ and $\zeta$. They can be expressed as
\cite{Leumer2020,Leumer2021,Leumer-2023}   
\begin{equation}\label{sympol}
\alpha(n)=Ae^{i\phi_1n}+Be^{-i\phi_1n}+Ce^{i\phi_2n}+De^{-i\phi_2n},
\end{equation}
where
\begin{equation}\label{eqdeftheta}
  \cos(\phi_\nu)=s_\nu/2, \quad \nu=1,2,
\end{equation}
\begin{equation}\label{eqfors}
  s_\nu=\frac{\eta\pm\sqrt{\eta^2-4(\zeta-2)}}{2},
\end{equation}
and provided $s_\nu^2\ne4$ and $s_1\ne s_2$.

The coefficients $A,B,C$ and $D$, are given by the initial values
$\alpha(j)$. In Ref.~\cite{Leumer-2023}, the values $j=-2,-1,0,1$, are   
used as seminal values while in this work we use $j=3,4,5,6$, since we
can always define $\alpha(n+5)=\xi(n)$, with both $\alpha$ and  $\xi$,
satisfiying Eq.~(\ref{symrecursion}). 

Notice that by substituting the ansatz, $\alpha(n)\propto w^n$, into
Eq.~(\ref{symrecursion}), we arrive at the characteristic equation, (see
Eq.~(\ref{charm})), with $\eta=\tau$, 
\begin{equation}\label{symcharpoly}
  w^4-\eta w^3+\zeta w^2 -\eta w+1=0.
\end{equation}
Hence, the eigenvalues of $\mathbf T_c$, are 
\begin{equation}\label{Meigenvalues}
  w_\mu=e^{\pm i\phi_1},e^{\pm i\phi_2}, \quad \mu=1,...,4,
\end{equation}
where $\phi_1$ and $\phi_2$ can be complex quantities. They
correspond to two waves propagating in both directions $\pm\hat z$
with phases $\phi_1$ and $\phi_2$, respectively.

We consider two examples of binary periodic optical system of type I
only, we will investigate type II binary cells in a future work. The
first one has a unit cell composed of two birefringent uniaxial media
and the second has a unit cell composed by two isotropic media subject
to an external magnetic field oriented along the $z$-direction,
exhibiting Faraday effect. For simplicity, we consider lossless
media. Under these assumptions, it turns out that $\eta$ and $\zeta$
are real valued \cite{Torres-2024} for both systems, then $s_\nu$,
$\nu=1,2$, are also real for $\eta^2\ge4(\zeta-2)$, and become
complex if $\eta^2<4(\zeta-2)$. Notice on the one hand, that in the
regime of real $s_\nu$, if $|s_\nu|\le 2$, then $\phi_\nu$ is a real
quantity and describes propagating oscillations in both directions
$\pm\hat z$; and if $|s_\nu|>2$, then $\phi_\nu$ is pure imaginary and
describes exponentially decaying or increasing oscillations.  
On the other hand, in the regime of complex values of $s_\nu$, we
obtain from Eq.~(\ref{eqfors}), $s_2=s_1^*$, therefore
$\phi_2=\pm\phi_1^*$. Also, according to Eq.~({\ref{Meigenvalues}),
both exponentials, $e^{\pm i\phi_2}$, are eigenvalues of
$\mathbf T_c$, therefore we can arbitrarily set $\phi_2=\phi_1^*$.   

\begin{figure}
\centering
\includegraphics[width=\columnwidth]{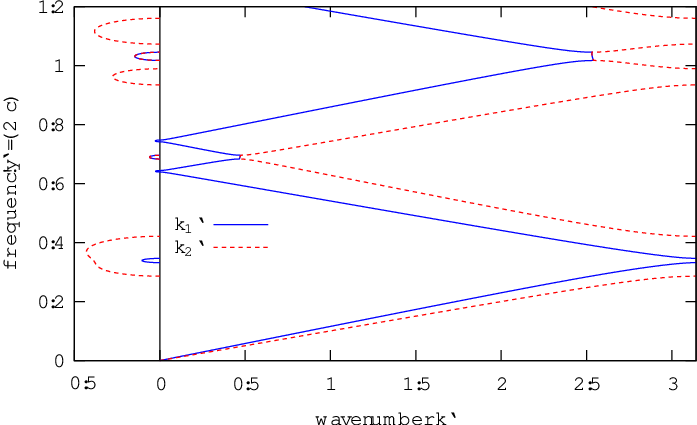}
\caption{Dispersion relation for a unit cell composed by two
  birefringent uniaxial media with their optical axis perpendicular to
  the $z$-axis, with ordinary refraction indexes, $n_o^{(1)}= 1.6$ and
  $n_o^{(2)}= 1.1$, and extraordinary refraction indexes $n_e^{(1)}=
  1.9$ and $n_e^{(2)}= 1.4$. The corresponding filling fractions are
  $f_1 = 0.4$ and $f_2 = 0.6$, respectively. We have assumed both the
  isotropic ambient and the substrate to have the same refraction
  index, $n_0=n_2=1$ and for simplicity, normal incidence is assumed,
  $\theta_0=0$. We have set the angles of the optical axes with
  respect to $x$-axis, $\psi_1$ and $\psi_2$, of the first and
  second layer,  $\psi_1=0$ and $\psi_2=\pi/4$, respectively, to
  allow the so-called mode coupling.  
}          
\label{bloc-zoom}
\end{figure}

In Figure \ref{bloc-zoom}, we show the dispersion relationship,
$\phi_\nu=k_\nu\ell$, $\nu=1,2$, where $\ell$, is the unit cell     
length, $\ell=\ell_1+\ell_2$, calculated from Eqs.~(\ref{eqdeftheta}),
(\ref{eqfors}) and (\ref{Meigenvalues}), for a binary cell consisting
of birefringent uniaxial media, with their optical axis perpendicular
to the $z$-axis. By convention, the imaginary part of $k_\nu \ell$ is
represented by negative abscissa values while the real part is
represented by positive ones. 
The interval of frequencies where $|s_\nu|\le 2$, either for $\nu=1$
or $\nu=2$ or both, defines the bands observed in Figure
\ref{bloc-zoom} ($k_\nu\ell>0$). 
Similarly, the interval of frequencies where $|s_\nu|> 2$, for both
$\nu=1,2$, defines the gap observed in Figure~\ref{bloc-zoom}
($k_\nu\ell<0$) around the value $\omega'=0.38$, where $\omega'$ is
the normalized frequency $\omega'=\frac{\omega\ell}{2\pi c}$.  
Around $\omega'=0.7$ and just above $\omega'=1$, both branches shown
in Figure~\ref{bloc-zoom}, overlap. Furthermore, the
corresponding wavenumbers $k_1$ and $k_2$ are complex, which
corresponds to attenuating or increasing propagating waves, and seem
to be degenerate, i. e. $k_1=k_2$. This scenario corresponds to the
regime of complex $s_\nu$ and according to the discussion in the above 
paragraph, $k_2=k_1^*$. Thus, despite the imaginary part of both
$k_1$ and $k_2$ have opposite signs, they are plotted with
the same sign in Figure 2. From Eq.~(\ref{eqfors}), the edges of these
complex bands, occur when $\eta^2=4(\zeta-2)$, clearly observable in
Figure~\ref{bloc-zoom}, where the wavenumbers also become
degenerate. The complex band around $\omega'=0.7$, connects two
typical bands, as well as does the complex band just above
$\omega'=1$. Finally, notice that as a thumb rule, when one of the
wavenumbers has an imaginary part in a particular interval of
frequency, the maximum value of its imaginary part is proportional to
the size of the interval of frequency. This thumb rule can be helpful
in understanding the partial transmission, as we will see in the next
section.     

\begin{figure}
\centering
\includegraphics[width=\columnwidth]{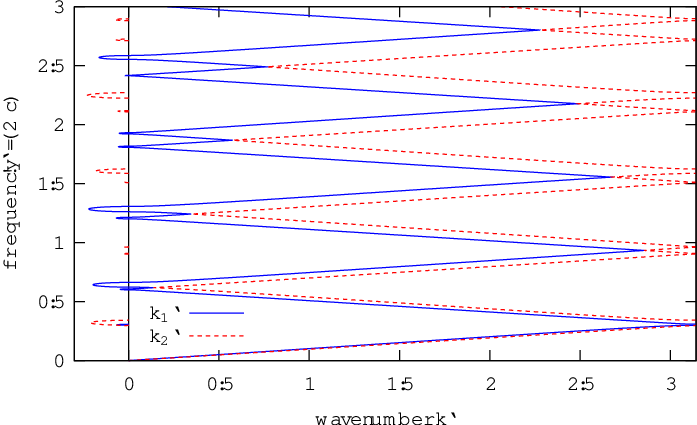}
\caption{Dispersion relation for a unit cell composed by two
  isotropic media that exhibit the Faraday effect when an external
  magnetic field is applied along the $z$-axis, with Faraday rotation
  parameters, $\gamma_1/\epsilon_v=0.36$ and
  $\gamma_2/\epsilon_v=0.001$, respectively. The corresponding
  refraction indexes of the first and second layer are, $n_o^{(1)}=
  1.47$ and  $n_o^{(2)}= 1.7$, with filling fractions
  $f_1 = 0.4$  and $f_2 = 0.6$, respectively. Again, we have assumed
  both the isotropic ambient and the substrate to have the same
  refraction index, $n_0=n_2=1$, and for simplicity, normal incidence
  is assumed, $\theta_0=0$.}          
\label{blochfar}
\end{figure}
 
In Figure~\ref{blochfar}, we show the dispersion relation
for our second example of binary cell, consisting of istropic media,
that exhibit the Faraday effect when an external magnetic field is
applied along the $z$-axis. A similar structure is observed as in our
first example. However, a closer look reveals the
absence of complex bands for the set of parameters chosen and for the
range of frequencies shown. Notice while in our first example, the
complex bands connected two typical bands, now these typical bands are
just joint at frequencies where $\eta^2=4(\zeta-2)$, i. e. the
wavenumbers become degenerate, $k_1=k_2$, at these frequencies.  

\section{Reflectance and transmittance}

We can use the closed forms for the symmetric Tetranacci
polynomials in Eq.~(\ref{symrecursion})
\cite{Leumer2020,Leumer2021,Leumer-2023},
\begin{equation}\label{closedformxi}
  \begin{array}{cc}
  \xi(n)=&\frac{1}{s_1-s_2}\left[\varphi_2(n)
    \left(\xi_{-2}-s_1\xi_{-1}+\xi_0\right) 
    -\varphi_1(n)\left(\xi_{-2}-s_2\xi_{-1}+\xi_0\right)\right.\\
    &\left.+\varphi_1(n+1)\left(\xi_{-1}-s_2\xi_0+\xi_1\right)
    -\varphi_2(n+1)\left(\xi_{-1}-s_1\xi_0+\xi_1\right)\right],
  \end{array}
\end{equation}
without degeneration, $s_1\ne s_2$, to calculate any
power of the transfer matrix $\mathbf T_c$, where we have employed the
notation of Ref.~\cite{Leumer-2023}, with
  $\xi(j)=\xi_j$, $j=-2,-1,0,1$.
$\varphi_\nu(n)$, $\nu=1,2,$ denotes the generalized Fibonacci
polynomial \cite{Webb,Hoggatt,Pashev}, defined by,
\begin{equation}\label{defphi}
  \varphi_\nu(n+1)=s_\nu\varphi_\nu(n)-\varphi_\nu(n-1), \quad n \in
  \mathbb Z, 
  \end{equation}
with $s_\nu$ given by Eq.~(\ref{eqfors}) and initial values
$\varphi_\nu(0)=0$, $\varphi_\nu(1)=1$.
By defining $\alpha(n+5)=\xi(n)$, we obtain $\alpha(3)=\xi_{-2}$,
$\alpha(4)=\xi_{-1}$, $\alpha(5)=\xi_0$ and $\alpha(6)=\xi_1$. Then we
can substitute the values $n=N-6$, $n=N-7$, $n=N-8$, and $n=N-9$, in
Eq.~(\ref{closedformxi}), to obtain $\alpha(n)$ for $n=N-1,N-2,N-3$
and $N-4$. The susbstitution of these $\alpha$-values in
Eq.~(\ref{beta})-(\ref{delta}) yields the coefficients $\beta(N-2)$,
$\gamma(N-3)$ and $\delta(N-4)$, and therefore, the $N$-power of
$\mathbf T_c$. We refer to the method that uses the closed forms of TP
of Eq.~(\ref{closedformxi}) as Method~B. Finally, we can calculate    
the transmittance or reflectance as usual \cite{Azzam}, from the
relation, 
\begin{equation} \label{film}
  \Psi_t=\mathbf T_c^N(\Psi_i+\Psi_r)
\end{equation}
where,
\begin{equation}\label{irt}
  \begin{array}{rcl}
    \Psi_i&=&\left(E_{ip}\cos\theta_0,\frac{E_{ip}}{Z_0},E_{is},\frac{E_{is}}{Z_0}\cos\theta_0\right)^T,\\
    \Psi_r&=&\left(-E_{rp}\cos\theta_0,\frac{E_{rp}}{Z_0},E_{rs},-\frac{E_{rs}}{Z_0}\cos\theta_0\right)^T,\\
    \Psi_t&=&\left(E_{tp}\cos\theta_2,\frac{E_{tp}}{Z_2},E_{ts},\frac{E_{ts}}{Z_2}\cos\theta_0\right)^T,
  \end{array}
\end{equation}
where $E_{\mu\nu}$, denotes the amplitude of the incident ($\mu=i$),
reflected ($\mu=r$), and transmitted electric field for
$\nu$-polarization, $\nu=p,s$. $\theta_2$ is the transmission angle
towards the substrate, $Z_0$ and $Z_2$, are the isotropic ambient and
substrate impedances, respectively. $T$ means transpose.      
The substitution of Eqs.~(\ref{irt}) into Eq.~(\ref{film}) yields
four linear equations in the six components, $E_{\mu\nu}$,
($\mu=i,r,t$; $\nu=s,p$). $E_{ts}$ can be eliminated from two of these
equations while $E_{tp}$ can in turn be eliminated from the remaining
two equations. This procedure yields two algebraic linear equations
for the incident and reflected fields alone, which can be written as, 
\begin{equation}\label{matrizR}
\bm E_r=\mathbf r \bm E_i,
\end{equation}
where $\bm E_\mu=(E_{\mu p},E_{\mu s})^T$, $\mu=i,r,t$ and $\mathbf r$
is the $2\times2$ complex-amplitude reflection matrix
\begin{equation}
  \mathbf r=\left(\begin{array}{cc}
    r_{pp}&r_{ps}\\
    r_{sp}&r_{ss}
  \end{array}\right),
\end{equation}
and is given by
\begin{equation}\label{rfilm}
  \mathbf r\!=\!(\mathcal P_r^s\mathcal Q_r^p- \mathcal
P_r^p\mathcal Q_r^s)^{-1}\left(\!\begin{array}{cc} 
\mathcal P_i^p\mathcal Q_r^s-\mathcal P_r^s\mathcal Q_i^p &\mathcal
P_i^s\mathcal Q_r^s-\mathcal 
P_r^s\mathcal Q_i^s\\ 
\mathcal P_r^p\mathcal Q_i^p-\mathcal P_i^p\mathcal Q_r^p&\mathcal
P_r^p\mathcal Q_i^s-\mathcal
P_i^s\mathcal Q_r^p\\ 
\end{array}\!\right)
\end{equation}  
where
\begin{equation}\label{ab}
\begin{array}{ccl}
\mathcal P_\mu^p&=&\pm\cos\theta_0(T_{11}-T_{21}Z_2\cos\theta_2)
+(T_{12}- T_{22}Z_2\cos\theta_2)/Z_0,\\  
\mathcal P_\mu^s&=&\pm\frac{\cos\theta_0}{Z_0}(T_{14}-
T_{24}Z_2\cos\theta_2)+T_{13}-T_{23}Z_2\cos\theta_2,\\  
\mathcal Q_\mu^p&=&\pm\cos\theta_0(T_{31}\frac{\cos\theta_2}{Z_2}-
T_{41})+ (T_{32}\frac{\cos\theta_2}{Z_2}-T_{42})/Z_0,\\ 
\mathcal Q_\mu^s&=&\pm\frac{\cos\theta_0}{Z_0}
(T_{34}\frac{\cos\theta_2}{Z_2}-T_{44})+
T_{33}\frac{\cos\theta_2}{Z_2}-T_{43},     
\end{array}
\end{equation}
$T_{mn}$ are the matrix elements of the transfer matrix $\mathbf
T_c^N$. In the above equations, the  $\mu=i$ and $\mu=r$ values on the
left correspond to the upper (+) and lower (-) signs on the right,
respectively. Finally, the reflectance, $R_\mu$, for incident
$\mu$-polarized light is given by
\begin{equation}
R_\mu=|r_{p\mu}|^2+|r_{s\mu}|^2,\quad \mu=p,s.
\end{equation}
Substitution of $\mathbf T_c^N$ into this equation yield a closed
form for the reflectance. Similarly, by substituting
Eq.~(\ref{matrizR}) into the second and third component of $\Psi_t$ of
Eq.~(\ref{film}),  the reflected field components are eliminated,
yielding two linear equations for the transmitted and incident
fields $\mathbf E_t$ and $\mathbf E_i$ only,
\begin{equation}\label{matrizt}
\bm E_t=\mathbf t \bm E_i,
\end{equation}
where $\mathbf t$ is the $2\times2$ complex-amplitude transmission
matrix 
\begin{equation}
  \mathbf t=\left(\begin{array}{cc}
    t_{pp}&t_{ps}\\
    t_{sp}&t_{ss}
  \end{array}\right),
\end{equation}
where
\begin{equation}\label{transmission}
\begin{array}{ccl}
  t_{pp}&=&Z_2\left[(T_{21}\cos\theta_0+\frac{T_{22}}{Z_0})(1-r_{pp})+
    r_{sp}(T_{23}-T_{24}\frac{\cos\theta_0}{Z_0})\right],\\
  t_{ps}&=&Z_2\left[T_{23}+T_{24}\frac{\cos\theta_0}{Z_0}
    +r_{ss}(T_{23}-T_{24}\frac{\cos\theta_0}{Z_0})
    +r_{ps}(\frac{T_{22}}{Z_0}-T_{21}\cos\theta_0)\right],\\
  t_{sp}&=&T_{31}\cos\theta_0+\frac{T_{32}}{Z_0}
  +r_{pp}(-T_{31}\cos\theta_0+\frac{T_{32}}{Z_0})
  +r_{sp}(T_{33}-T_{34}\frac{\cos\theta_0}{Z_0}),\\
  t_{ss}&=&T_{33}+T_{34}\frac{\cos\theta_0}{Z_0}
  +r_{ss}(T_{33}-T_{34}\frac{\cos\theta_0}{Z_0})
  +r_{ps}(\frac{T_{32}}{Z_0}-T_{31}\cos\theta_0).\\
\end{array}
\end{equation}
Therefore, the transmittance, $T_\mu$, for incident $\mu$-polarized 
light is given by
\begin{equation}
  T_\mu=\frac{Z_0\cos\theta_2}{Z_2\cos\theta_0}\left(|t_{p\mu}|^2+
  |t_{s\mu}|^2\right), \quad \mu=p,s. 
\end{equation}

For $\eta$ and $\zeta$ real \cite{Torres-2024}, the condition
$s_1=s_2$, occurs at the edges of the complex bands, i.e. when
$\eta^2=4(\zeta-2)$. Thus, the corresponding expressions for $\xi(n)$
in the degenerate case, $s_1=s_2$, with either $s_1^2\ne4$
  or $s_1^2=4$, are required. They can be obtained \cite{Torres-2024}
  from preliminary steps shown in \cite{Leumer-2023}, 
  and we reproduce them here: If $s_1=s_2$ but $s_1^2\ne4$, then  
\begin{equation}\label{xis1eqs2}
  \begin{array}{cl}
    \xi(n)=&\frac{1}{s_1^2-4}\left[\varphi_1(n+2)\left(n\xi_1
      -3(n+1)s_1\xi_0\right)\right.\\   
      &+\varphi_1(n+1)\left((1-n)\xi_{-2}+2(n+2)(s_1^2-1)\xi_0\right)\\
      &+\varphi_1(n)\left(-(n+2)\xi_1+2(n-1)(s_1^2-1)\xi_{-1}\right)\\
      &\left.+\varphi_1(n-1)\left((n+1)\xi_{-2}-3ns_1\xi_{-1}\right)
      \right],
  \end{array}
\end{equation}
if $s_1=s_2$ and $s_1^2=4$, then 
\begin{equation}\label{xis12eq4}
  \begin{array}{cc}
    \xi(n)=&\frac{s_1}{12}\left[(n+1)(n+3)\xi_0\varphi_1(n+2) 
      +n(n+2)(\xi_1-2s_1\xi_0)\varphi_1(n+1)\right.\\
      &\left.+(n^2-1)(2s_1\xi_{-1}-\xi_{-2})\varphi_1(n)
      +n(2-n)\xi_{-1}\varphi_1(n-1)\right].
  \end{array}
\end{equation}
These expressions allow to treat the degeneracy of the $\mathbf T_c$
eigenvalues \cite{Laufer-1997}, that was not possible to do in
Ref.~\cite{Lin-1969}. 

\begin{figure}
\centering
\includegraphics[width=\columnwidth]{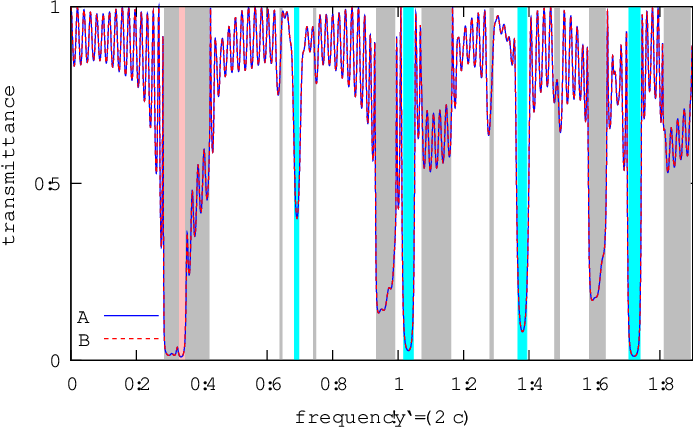}
\caption{Transmittance for normally incident $p$-polarized light on the
  finite periodic optical system with unit cell described
  in Figure 1 and consisting of $N=16$ cells. The results obtained
  with both methods, $A$ and $B$, are shown. They are
  indistinguishable. 
}    
\label{methods}
\end{figure}

In Figure~\ref{methods}, we compare both methods, $A$ and $B$, i.e. the  
iteration of Eq.~(\ref{symrecursion}) up to $n=N-1$,
from its seminal values and the one that uses the closed forms of
Eqs.~(\ref{closedformxi})-(\ref{defphi}), and
(\ref{xis1eqs2})-(\ref{xis12eq4}), to calculate the 
transmittance for normally incident $p$-polarized light on the finite
periodic optical system with unit cell described in
Figure~\ref{bloc-zoom} and consisting of $N=16$ cells.
We can observe that both results are indistinguishable. Notice that
the range of frequencies where the transmittance is attenuated,
corresponds with the gaps for the infinite system (cyan strips and
shown in Figure~\ref{bloc-zoom}), where both wavenumbers, $k_1$ and
$k_2$, are complex. The first gap (pink strip) is different in 
nature and indicates the interval of frequency where both, $k_1$ 
and $k_2$, are pure imaginary, corresponding to exponentially decaying 
or increasing optical field oscillations. Gray strips indicate the
interval of frequencies where one of the wavenumbers is imaginary pure
and the other is real. The rest of the intervals of frequency
correspond to real wavenumbers, i.e. propagating waves. The first
obvious observation is that when we have propagating waves, they can
interfer constructively or destructively, in our case partially
destructive, as it can be observed in the oscillations of the
transmittance just below the level of maximum transmission.
Also observe the relatively wide intervals of frequency indicated by
gray areas around a gap, indicative of a relatively large imaginary
pure wavenumber, where the transmittance drops drastically
for frequencies below the gap compared with the transmittance for
frequencies above the gap. This effect might indicative of a
destructive/constructive interference between the different
exponentially decaying or increasing optical field oscillations, and
can be studied only in the regime of a finite periodic optical system.   
This is one of the main results of this work. In contrast, this effect
fades in the relatively narrow intervals of frequency indicated by
gray areas around a gap, possibly due to the interference process,
just discussed, become negligible.

\begin{figure}
\centering
\includegraphics[width=\columnwidth]{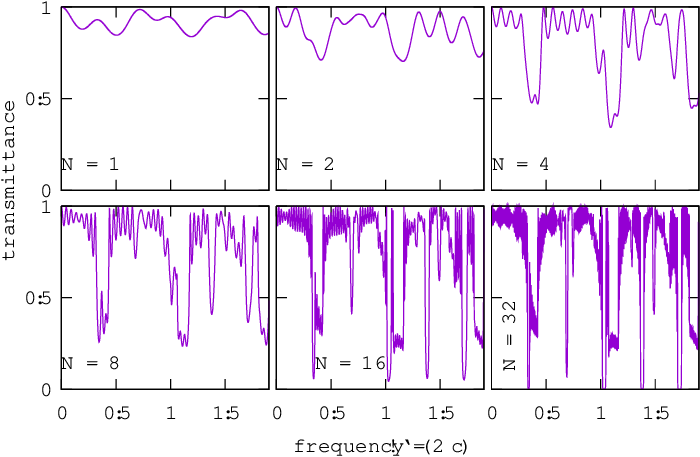}
\caption{(Color online) Transmittance of the same system of Figure~1
  as function of the number of cells and for normally incident
  $s$-polarized light.  
}    
\label{spol}
\end{figure}

\begin{figure}
\centering
\includegraphics[width=\columnwidth]{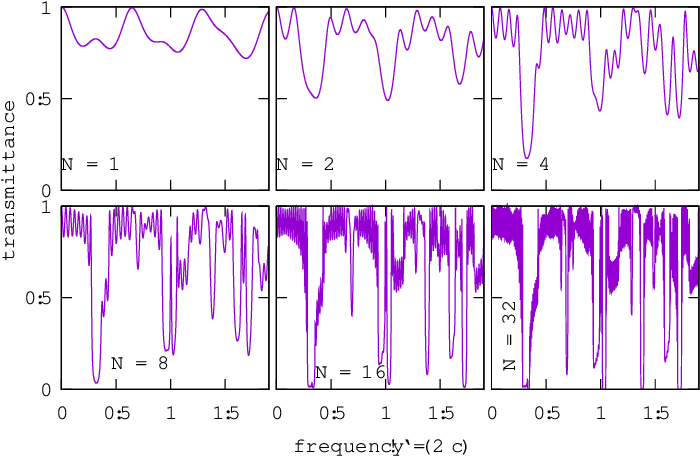}
\caption{Transmittance of the same system of Figure~1, as function of
  the number of cells for normally incident $p$-polarized light.
}    
\label{ppol}
\end{figure}

In Figure~\ref{spol}, we show the transmittance for the same periodic
system as in Figure 1, as a function of the number of cells for
normally incident $s$-polarized light. In Figure~\ref{ppol}, we show
the corresponding result for normally incident $p$-polarized light.  
Although the band formation is evident, notice however that the drop
in the transmittance around the emerging gaps, as a function of the
number of cells, has larger velocity for incident $p$-polarization as
the number of cells increases. To analize this subtle effect, it is
important to have a reliable method to precisely calculate the
spectra. For instance, notice that around $\omega'=1.6$ and for
$N=32$, the transmittance drops significatively for normally incident
$p$-polarized light while for $s$-polarized light transmittance is
still relatively high. This might be due to the above mentioned effect
of destructive/constructive interference between the different
exponentially decaying or increasing field oscillations, and 
can be thoroughly investigated with our formalism.

\begin{figure}
\centering
\includegraphics[width=\columnwidth]{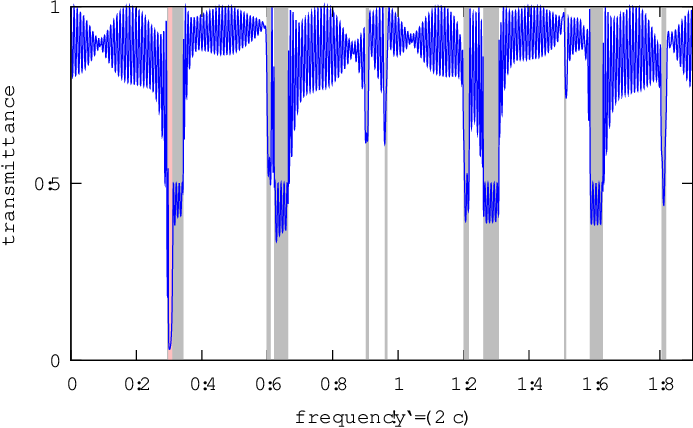}
\caption{(Color online) Transmittance for either normally incident
  $s$ or $p$-polarized light on the finite periodic optical system
  with unit cell described in Figure 2 and consisting of $N=32$
  cells. 
}    
\label{far-bandsgap-p}
\end{figure}

Finally, in Figure~\ref{far-bandsgap-p}, we show transmittance for the
same periodic system as in Figure 2 and consisting of $N=32$ cells,
for either normally incident $s$ or $p$-polarized light. Notice 
that only for normal incidence and isotropic media, makes no sense
to distinguish between polarizations of light. Therefore both
polarizations yield the same result, as is indeed the case. We
can observe that in the relatively wide gray intervals of frequency,
the transmittance drops significatively but not as low as in the gap
(pink area).

\section{Conclusions}
\label{conclusions}
In this work, the transmittance of a finite periodic anisotropic
optical system, is obtained in a closed form. The only restrictions on
the $4\times4$ differential propagation matrix of each layer composing
the binary unit cell, $\mathbb\Delta$, are $a)$~that its eigenvalues are
of the form $\pm~p_1$, $\pm~p_2$, with $p_1\ne p_2$, and $b)$ its
off-diagonal $2\times2$-block matrices possess the same symmetric
structure in both layers. Then, for an arbitrary number of cells, $N$,
the problem is reduced to calculating the $N$-th power of the transfer
matrix of a unit cell. The Cayley-Hamilton theorem is invoked and a set of
recursive relations are obtained, that leads to the recursive relation
defining the Tetranacci Polynomials. Such recursive relations are
solvable for any invertible $4\times4$-matrix and provide an algorithm
to compute the $N$-power of the transfer matrix avoiding the
multiplication of $N$ matrices. Furthermore, by using the recently closed
expressions for symmetric TP, the transmittance of a finite periodic
stratified and anisotropic optical system, fulfilling the above
mentioned conditions, and for an arbitrary number of unit cells, is
obtained in a closed form. The results for two birefringent uniaxial
media with their optical axis perpendicular to the $z$-axis and for
two isotropic media subject to an external magnetic field oriented
along the $z$-axis, exhibiting Faraday effect, are compared for both
methods, finding them indistinguishable. Finally, lossy media with
magnetic anisotropy and/or optical activity can be treated with our
formalism. 

\section{Acknowledgements}
ADA and JA were supported by CONACyT under project number
A1-S-23120. JCTG acknowledge financial support from CONACyT, Mexico
via posdoctoral grant from the program Estancias Posdoctorales por
México.  

\section{Declaration of interests}
The authors declare that they have no known competing financial
interests or personal relationships that could have appeared to
influence the work reported in this paper.

\section{Data availability}
No datasets were analyzed or generated during the course of the
current study.  

\section{Appendix A: The eigenvector matrix}
Let us decompose the $\mathbb\Delta$-matrix into $2\times2$ matrices
as follows,
\begin{equation}
  \mathbb\Delta=\left( \begin{array}{cc}
    \mathbb\Delta_{11}&    \mathbb\Delta_{12}\\
    \mathbb\Delta_{21}&    \mathbb\Delta_{22}
    \end{array}\right).
\end{equation}
The conditions for an even characteristic polynomial,
Tr~$\mathbb\Delta=0$ and Tr~$\mathbb\Delta^3=0$
\cite{BarutO-1994,BarutSU-1994,Torres-2024}, can 
  be fulfilled by the following sufficient conditions, 
\begin{equation}\label{traza-delta}
  \mbox{Tr }\mathbb\Delta_{11}= \mbox{Tr }\mathbb\Delta_{22}=0,
\end{equation}
and
\begin{equation}\label{traza-deltacubica}
  \mbox{Tr
  }\left(\mathbb\Delta_{11}\mathbb\Delta_{12}\mathbb\Delta_{21}
  +\mathbb\Delta_{22}\mathbb\Delta_{21}\mathbb\Delta_{12}\right)=0,
\end{equation}
respectively, where we have used the first condition in the second.
The corresponding non-degeneracy condition,
  $p_1\ne~p_2$, derives in 
two additional conditions, $|\mathbb\Delta|\ne0$, and  
$(\mbox{Tr }\mathbb\Delta^2)^2\ne16\mbox{ det~}\mathbb\Delta$. In terms
of $2\times2$-matrices, the latter reads as
\begin{equation}\label{non-deg}
  \left(\mbox{Tr~}\mathbb\Delta_{11}^2+\mbox{Tr~}\mathbb\Delta_{22}^2 
 +2\mbox{ Tr~}\mathbb\Delta_{12}\mathbb\Delta_{21}\right)^2\ne
 16\mbox{ det~}\mathbb\Delta. 
\end{equation}

An immediate option to fulfill Eq.~(\ref{traza-deltacubica}) is that
both matrices, $\mathbb\Delta_{11}$ and $\mathbb\Delta_{22}$, besides
being traceless according to the condition of Eq.~(\ref{traza-delta}),
are null matrices. We discard this option, since these matrices are
related to the diagonal of electrical permitivity and magnetic
permeability tensors, which is very unlikely they are null
\cite{Berreman-1972,Azzam}. A second option is the  case where one or
both matrices, $\mathbb\Delta_{12}$ and $\mathbb\Delta_{21}$, are
null. Notice that in this case we may find the widely studied
isotropic case, in which the electromagnetic polarizations decouple
and we will not consider here. Then, if one or both matrices, 
$\mathbb\Delta_{12}$ and $\mathbb\Delta_{21}$, are null and the medium
is not isotropic, the non-degeneracy condition of Eq.~(\ref{non-deg})
may not be fulfilled. In this case,
\begin{equation}
  \mbox{det }\mathbb\Delta= \frac{1}{4}\mbox{ Tr~}\mathbb\Delta_{11}^2
  \mbox{ Tr~}\mathbb\Delta_{22}^2,
\end{equation}
and the condition of Eq.~(\ref{non-deg}) becomes,
\begin{equation}\label{non-deg2}
  \left(\mbox{Tr~}\mathbb\Delta_{11}^2+\mbox{Tr~}\mathbb\Delta_{22}^2 
 \right)^2\ne  4 \mbox{ Tr~}\mathbb\Delta_{11}^2
  \mbox{ Tr~}\mathbb\Delta_{22}^2,
\end{equation}
which evidently is not fulfilled if $\mbox{ Tr~}\mathbb\Delta_{11}^2
  =\mbox{ Tr~}\mathbb\Delta_{22}^2$. Therefore, if one or both
  matrices, $\mathbb\Delta_{12}$ and $\mathbb\Delta_{21}$, are 
null, then we must require in addition,
\begin{equation}\label{extracond}
  \mbox{ Tr~}\mathbb\Delta_{11}^2\ne\mbox{ Tr~}\mathbb\Delta_{22}^2,
\end{equation}
to guarantee the non-degeneracy condition. We can include this option,
along with the extra condition of Eq.~(\ref{extracond}), as a special
case of a third option, where both matrices,
$\mathbb\Delta_{12}\mathbb\Delta_{21}$ 
and $\mathbb\Delta_{21}\mathbb\Delta_{12}$, have zero off-diagonal
elements, since their multiplication with a traceless matrix
(Eq.~(\ref{traza-delta})) yield a traceless matrix.
This option in turn implies that both matrices, $\mathbb\Delta_{12}$
and $\mathbb\Delta_{21}$, either have zero diagonal elements or both 
have zero off-diagonal elements. Then, we distinguish these two types 
of matrices as I and II, respectively. According to
Eq.~(\ref{traza-delta}), then $\mathbb\Delta_{11}$ and
$\mathbb\Delta_{22}$, are of type I. As another
  example, the matrices $\mathbb\Delta_{12}$ and $\mathbb\Delta_{21}$, 
  appearing in free oscillations of Timoshenko-Ehrenfest beams, are of  
  type II \cite{Torres-2024}. If one of the matrices,  
  $\mathbb\Delta_{12}$ and $\mathbb\Delta_{21}$, is null, then it can be
  considered of the same type as the other non-null matrix. If both
  matrices, $\mathbb\Delta_{12}$ and $\mathbb\Delta_{21}$, are null, then
  they can be considered of any type, which is convenient to match the
  type of the other layer in the binary cell. In the latter two special
  cases, the medium has to fulfill in addition the non-degeneracy
  condition of Eq.~(\ref{extracond}). 

Let us now see the implications on the eigenvector matrix 
$\mathbb V=(\vec V_1,\vec V_2,\vec V_3,\vec V_4)$.
For each eigenvalue, $p_j$ (or $k_j$), $j=1,2,3,4$, the corresponding    
eigenvector $\vec V_j$ is obtained from Eq.~(\ref{eigenvalor}) and
therefore, the matrix $\mathbb\Delta$ satisfies
\begin{equation}\label{delta-eigen}
\mathbb \Delta=\mathbb{V}\mathbb{Y}\mathbb{V}^{-1},
\end{equation}
where $\mathbb{Y}=\mbox{ diag }(p_1,p_2,-p_1,-p_2)$, $p_1\ne~p_2$.
We emphasize Eq.~(\ref{delta-eigen}) is valid only if
  $p_1\ne~p_2$, 
otherwise $\mathbb\Delta$ might not have four independent
eigenvectors. However, as mentioned in Section~4, from a practical
point of view, we can recover the case of an isotropic layer, in which
$p_1=p_2$, to any degree of approximation by taking the anisotropy
arbitrarily small, such that the above expression is still valid to
avoid discontinuous solutions \cite{Nikolai,Xu-2000,Zhang-2015}. 
Let us rewrite Eq.~(\ref{delta-eigen}) in terms of $2\times2$
matrices. First, we define  
\begin{equation}
  \mathbb V=\left( \begin{array}{cc}
    \mathbb V_{11}&    \mathbb V_{12}\\
    \mathbb V_{21}&    \mathbb V_{22}
    \end{array}\right),
\end{equation}
and
\begin{equation}
  \mathbb Y=\left( \begin{array}{cc}
    \mathbb\upsilon&    \mathbb 0\\
    \mathbb0& -\mathbb\upsilon
    \end{array}\right),
\end{equation}
where $\mathbb{0}$ stands for the null $2\times2$-matrix and
$\mathbb \upsilon$ is the diagonal $2\times2$-matrix, $\mathbb
\upsilon=\mbox{ diag }(p_1,p_2)$. Also, $\mathbb V_{\mu\nu}$,
$\mu,\nu=1,2$, are $2\times2$-matrices.
The substitution into
Eq.~(\ref{delta-eigen}) yields four $2\times2$-matrix equations,
\begin{eqnarray}
  \mathbb\Delta_{11}\mathbb V_{11}+\mathbb\Delta_{12}\mathbb V_{21}=
  \mathbb V_{11}\mathbb\upsilon,\label{v1}\\
  \mathbb\Delta_{21}\mathbb V_{11}+\mathbb\Delta_{22}\mathbb V_{21}=
  \mathbb V_{21}\mathbb\upsilon,\label{v2}\\
  \mathbb\Delta_{11}\mathbb V_{12}+\mathbb\Delta_{12}\mathbb V_{22}=
  -\mathbb V_{12}\mathbb\upsilon,\label{v3}\\
  \mathbb\Delta_{21}\mathbb V_{12}+\mathbb\Delta_{22}\mathbb V_{22}=
  -\mathbb V_{22}\mathbb\upsilon.\label{v4}
\end{eqnarray}
Notice that Eqs.~(\ref{v1}) and (\ref{v3}) are equivalent if a
relation exists between $\mathbb V_{11}$ and $\mathbb V_{12}$, and 
between $\mathbb V_{22}$ and $\mathbb V_{21}$, which
is an expected result, in the sense that if we know $\vec V_1$ 
and $\vec V_2$ eigenvectors, then we just need to replace $p_1$ by
$-p_1$ and $p_2$ by $-p_2$ to obtain the eigenvectors $\vec V_3$ and
$\vec V_4$. 
A similar statement applies to Eqs.~(\ref{v2}) and (\ref{v4}).

For matrices $\mathbb\Delta_{12}$ and $\mathbb\Delta_{21}$, of type I, 
i.e. with zero diagonal elements 
(as $\mathbb\Delta_{11}$ and $\mathbb\Delta_{22}$), we have
\begin{equation}\label{sigmaprop}
  \mathbb\sigma_z\mathbb\Delta_{\mu\nu}\mathbb\sigma_z=
  -\mathbb\Delta_{\mu\nu},\quad \mu,\nu=1,2, 
\end{equation}
where $\mathbb\sigma_z$ is the Pauli matrix,
\begin{equation}
  \mathbb\sigma_z=\left(\begin{array}{cc}
    1&0\\
    0&-1
  \end{array}
  \right).
\end{equation}
Therefore, Eq.~(\ref{v1}) can be rewritten as 
\begin{equation}
  -\mathbb\sigma_z\mathbb\Delta_{11}\mathbb\sigma_z\mathbb V_{11}-
  \mathbb\sigma_z\mathbb\Delta_{12}\mathbb\sigma_z\mathbb V_{21}=
  \mathbb\sigma_z^2\mathbb V_{11}\mathbb\upsilon, 
\end{equation}
where we have used $\sigma_z^2=\mathbf I$, and
$\mathbf I$ is the identity $2\times2$-matrix. 
Factorizing a $\mathbb\sigma_z$-matrix by the left we obtain,
\begin{equation}
  -\mathbb\Delta_{11}\mathbb\sigma_z\mathbb{V_{11}}-
  \mathbb\Delta_{12}\mathbb\sigma_z\mathbb{V_{21}}=
  \mathbb\sigma_z\mathbb{V_{11}}\mathbb{\upsilon}, 
\end{equation}
and comparing with Eq.~(\ref{v3}) yields
\begin{eqnarray}
  \mathbb V_{12}=a\mathbb\sigma_z\mathbb V_{11}\label{v11-v12}\\
  \mathbb V_{22}=a\mathbb\sigma_z\mathbb V_{21},\label{v21-v22}
\end{eqnarray}
where $a$ is non-zero constant.
By substituting Eqs.~(\ref{v11-v12})-(\ref{v21-v22}) into
Eq.~(\ref{v4}) and multiplying by the left by $\mathbb\sigma_z$
we obtain
\begin{equation}
  a\mathbb\sigma_z\mathbb\Delta_{21}\mathbb\sigma_z\mathbb V_{11}+a
  \mathbb\sigma_z\mathbb\Delta_{22}\mathbb\sigma_z\mathbb V_{21}=
  -a\mathbb\sigma_z^2\mathbb V_{21}\mathbb\upsilon, 
\end{equation}
by using Eq.~(\ref{sigmaprop}) again, this Eq. simplifies to
\begin{equation}
  -a\mathbb\Delta_{21}\mathbb V_{11}-a
  \mathbb\Delta_{22}\mathbb V_{21}=
  -a\mathbb V_{21}\mathbb\upsilon, 
\end{equation}
finally, by factorizing $-a$ we recover Eq.~(\ref{v2}).

For matrices $\mathbb\Delta_{12}$ and $\mathbb\Delta_{21}$, of type 
II, i.e. with zero off-diagonal elements, 
\begin{equation}\label{sigmaprop-2}
  \mathbb\sigma_z\mathbb\Delta_{\mu\nu}\mathbb\sigma_z=
  \left\{\begin{array}{c}
      -\mathbb\Delta_{\mu\nu},\quad \mbox{ for }\mu=\nu,\\
      \mathbb\Delta_{\mu\nu},\quad \mbox{ for } \mu\ne\nu,
      \end{array}\right.
\end{equation}
and by analyzing Eqs.~(\ref{v1})-(\ref{v4}) we obtain instead of
Eq.~(\ref{v21-v22}),    
\begin{equation}
  \mathbb{V_{22}}=-a\mathbb\sigma_z\mathbb{V_{21}}.
\end{equation}
Notice that the above results remain unchanged if one or both 
matrices, $\mathbb\Delta_{12}$ or $\mathbb\Delta_{21}$, are null.

The above procedure on $\mathbb{V}$, can be repeated on matrix
$\mathbb{V^{-1}}$, 
\begin{equation}
  \mathbb V^{-1}=\left( \begin{array}{cc}
    \mathbb U_{11}&    \mathbb U_{12}\\
    \mathbb U_{21}&    \mathbb U_{22}
    \end{array}\right),
\end{equation}
to yield
\begin{eqnarray}
  \mathbb{U_{21}}=a^{-1}\mathbb{U_{11}}\mathbb\sigma_z\label{u11-u21}\\
  \mathbb{U_{22}}=\pm a^{-1}\mathbb{U_{12}}\mathbb\sigma_z,\label{u12-u22}
\end{eqnarray}
where the plus or minus sign is taken for
$\mathbb\Delta_{12}$,~$\mathbb\Delta_{21}$-matrices of the type I or
II, respectively. The inverse of the constant $a$, is required
to satisfy the condition that $\mathbb V\mathbb V^{-1}=\mathbb
V^{-1}\mathbb V=\mathbf I$, where $\mathbf I$, is the identity
$4\times4$-matrix. 
Therefore, the $\mathbb{V}$ and $\mathbb{V^{-1}}$ matrices, can be
expressed in terms only of its two left and upper
$2\times2$-matrices, respectively,
\begin{equation}\label{vmatrix}
  \mathbb V=\left( \begin{array}{cc}
    \mathbb V_{11}&    a\mathbb\sigma_z\mathbb V_{11}\\
    \mathbb V_{21}&    \pm a\mathbb\sigma_z\mathbb V_{21}
    \end{array}\right),
\end{equation}
\begin{equation}\label{umatrix}
  \mathbb V^{-1}=\left( \begin{array}{cc}
    \mathbb U_{11}&\mathbb U_{12}\\
    a^{-1}\mathbb U_{11}\mathbb\sigma_z
    &  \pm a^{-1}\mathbb U_{12}\mathbb\sigma_z
    \end{array}\right).
\end{equation}
The result of Eq.~(\ref{vmatrix}), states that two of the $\mathbb
\Delta$-eigenvectors can be written in terms on the other two, 
which is not surprising as we stated above, since we simply have to
change the sign of $p_1$ and $p_2$ in the calculation of the
corresponding eigenvectors. Therefore, the constant $a$, enters in
play since eigenvectors are defined upto a constant. Hence, 
without lost of generality we can set $a=1$ from now on. 

\section{Appendix B: The symmetric condition $\eta=\tau$}
Let us now construct the matrix $\mathbb W=\mathbb V_1^{-1}\mathbb
V_2$, where the subindexes label the first or the second anisotropic
layer in the unitary cell, by using Eqs.~(\ref{vmatrix}) and
(\ref{umatrix}) with $a=1$, we obtain, 
\begin{equation}
  \mathbb W=\left(\begin{array}{cc}
    \mathbb \Omega+\mathbb\Lambda& \mathbb \Gamma+\iota_2\mathbb\Sigma\\
    \mathbb \Gamma+\iota_1\mathbb\Sigma& \mathbb
    \Omega+\iota_1\iota_2\mathbb\Lambda 
    \end{array}\right)
\end{equation}
where $\mathbb\Omega$, $\mathbb\Gamma$, $\mathbb\Lambda$ and
$\mathbb\Sigma$, are the $2\times2$-matrices,
\begin{equation}\label{Omegas}
  \begin{array}{c}
    \mathbb \Omega= \mathbb U_{11}^{(1)}\mathbb V_{11}^{(2)},\\
    \mathbb \Lambda= \mathbb U_{12}^{(1)}\mathbb V_{21}^{(2)},\\
    \mathbb \Gamma= \mathbb U_{11}^{(1)}\sigma_z\mathbb
    V_{11}^{(2)},\\
    \mathbb \Sigma= \mathbb U_{12}^{(1)}\sigma_z\mathbb V_{21}^{(2)},
  \end{array}
\end{equation}
and $\iota_j$, $j=1,2$, characterize the type of both of the off-diagonal
matrices $\mathbb\Delta_{12}$ and $\mathbb\Delta_{21}$ corresponding
to the $j$-th layer. Thus, $\iota_j=\pm1$ for type I and II,
respectively. In Eq.~(\ref{Omegas}), the superscripts label the first
or the second anisotropic layer in the unitary cell.

Notice $\mathbb W^{-1}=\mathbb V_2^{-1}\mathbb 
V_1$.  Therefore, the inverse of $\mathbb W$, $\mathbb W^{-1}$,
has also a similar form,
\begin{equation}
  \mathbb W^{-1}=\left(\begin{array}{cc}
    \mathbb \Omega'+\mathbb\Lambda'& \mathbb \Gamma'+\iota_1\mathbb\Sigma'\\
    \mathbb \Gamma'+\iota_2\mathbb\Sigma'& \mathbb
    \Omega'+\iota_1\iota_2\mathbb\Lambda' 
    \end{array}\right)
\end{equation}
where $\mathbb\Omega'$, $\mathbb\Gamma'$, $\mathbb\Lambda'$ and
$\mathbb\Sigma'$, are the $2\times2$-matrices,
\begin{equation}
  \begin{array}{c}
    \mathbb \Omega'= \mathbb U_{11}^{(2)}\mathbb V_{11}^{(1)},\\
    \mathbb \Lambda'= \mathbb U_{12}^{(2)}\mathbb V_{21}^{(1)},\\
    \mathbb \Gamma'= \mathbb U_{11}^{(2)}\sigma_z\mathbb
    V_{11}^{(1)},\\
    \mathbb \Sigma'= \mathbb U_{12}^{(2)}\sigma_z\mathbb V_{21}^{(1)}.
  \end{array}
\end{equation}

Let us consider the trace of the transfer matrix $\mathbf T_c$
\begin{equation}\label{traza}
  \mbox{Tr }\mathbf T_c = \mbox{Tr
  }\mathbb{V}_1^{-1}\mathbb{V}_2\mathbb
  \lambda_2(\ell_2)\mathbb{V}_2^{-1}\mathbb{V}_1\mathbb
  \lambda_1(\ell_1) =\mbox{ Tr }\mathbb W \lambda_2(\ell_2) \mathbb W^{-1}
  \lambda_1(\ell_1),    
\end{equation}
where $\mathbb\lambda_j(\ell_j)$ is the eigenvalue matrix of
Eq.~(\ref{eigenvalues}) but corresponding to the $j$-th layer in the
unit cell. Also, we have made use of Eqs.~(\ref{eqforT}) and
(\ref{tm}), and of the ciclic property of the trace operation. Now, by
rewriting $\mathbb\lambda(\ell_j)$, $j=1,2$  
\begin{equation}
  \lambda(\ell_j)=\left(\begin{array}{cc}
    \mathbb K_j& \mathbb{0}\\
    \mathbb{0}& \mathbb K_j^{-1}
    \end{array}\right)
\end{equation}
where $\mathbb{0}$, stands for the null $2\times2$-matrix, and
$\mathbb K_j$, is the diagonal $2\times2$-matrix $\mathbb
K_j=\mbox{ diag }(e^{ik_1^{(j)}\ell_j},e^{ik_2^{(j)}\ell_j})$, where
$k_1^{(j)}$ and $k_2^{(j)}$ are the eigenvalues of
Eq.~(\ref{eigenvalues}) corresponding to the $j$-th layer in the
unit cell. By expanding the matrix multiplication in Eq.~(\ref{traza})
in terms of $2\times2$-matrices, we obtain
\begin{equation}\label{trace}   
  \begin{array}{ll}
    \mbox{Tr }\mathbf T_c =& \mbox{Tr}\left[
    (\mathbb{\Omega}+\mathbb\Lambda)\mathbb{K}_2(\mathbb\Omega'+
    \mathbb\Lambda')\mathbb K_1 + 
    (\mathbb{\Gamma}+\iota_2\mathbb\Sigma)\mathbb{K}_2^{-1}(\mathbb
    \Gamma'+\iota_2\mathbb\Sigma')\mathbb K_1 \right.\\
  &\left.+(\mathbb{\Omega}+\iota_1\iota_2\mathbb\Lambda)\mathbb{K}_2^{-1}(\mathbb 
  \Omega'+\iota_1\iota_2\mathbb\Lambda')\mathbb K_1^{-1} 
  +(\mathbb\Gamma+\iota_1\mathbb\Sigma)\mathbb{K}_2(\mathbb\Gamma'+
  \iota_1\mathbb\Sigma')\mathbb K_1^{-1} \right].
    \end{array}
\end{equation}
Similarly,  
\begin{equation}
  \mbox{Tr }\mathbf T_c^{-1} = \mbox{ Tr }\mathbb W
  \mathbb\lambda_2^{-1}(\ell_2) \mathbb W^{-1}
  \mathbb\lambda_1^{-1}(\ell_1),     
\end{equation}
therefore, to obtain Tr $\mathbf T_c^{-1}$, we just have to replace
$\mathbb K_j$ by $\mathbb K_j^{-1}$ and viceversa in
Eq.~(\ref{trace}),
\begin{equation}\label{traceinv}   
  \begin{array}{ll}
    \mbox{Tr }\mathbf T_c^{-1} =& \mbox{Tr}\left[
    (\mathbb{\Omega}+\mathbb\Lambda)\mathbb{K}_2^{-1}(\mathbb\Omega'+
    \mathbb\Lambda')\mathbb K_1^{-1} + 
    (\mathbb{\Gamma}+\iota_2\mathbb\Sigma)\mathbb{K}_2(\mathbb
    \Gamma'+\iota_2\mathbb\Sigma')\mathbb K_1^{-1} \right.\\
  &\left.+(\mathbb{\Omega}+\iota_1\iota_2\mathbb\Lambda)\mathbb{K}_2(\mathbb 
  \Omega'+\iota_1\iota_2\mathbb\Lambda')\mathbb K_1 
  +(\mathbb\Gamma+\iota_1\mathbb\Sigma)\mathbb{K}_2^{-1}(\mathbb\Gamma'+
  \iota_1\mathbb\Sigma')\mathbb K_1 \right],
    \end{array}
\end{equation}
therefore, this operation leaves the rhs of Eq.~(\ref{trace}) the same
only if $\iota_1$ and $\iota_2$ have the same sign, i.e. if the
off-diagonal matrices $\mathbb\Delta_{12}$ and $\mathbb\Delta_{21}$ in
both layers are of the same type. Only in this case we have Tr
$\mathbf T_c$= Tr $\mathbf T_c^{-1}$. 

Finally, if the transfer matrix, $\mathbb T_c$, for a unitary cell
composed by two anisotropic layers satisfies Eq.~(\ref{charm}), then
its inverse, $\mathbf T_c^{-1}$, satisfies  
\begin{equation}\label{charminv}
  (\mathbf T_c^{-1})^4-\tau (\mathbf T_c^{-1})^3+\zeta (\mathbf T_c^{-1})^2
  -\eta \mathbf T_c^{-1} +\mathbf{I_4}=\mathbb{0}.
\end{equation}
Thus, the condition, $\eta=\tau$, implies Tr~$\mathbf T_c$=Tr~$\mathbf
T_c^{-1}$.

\section{Appendix C: Examples of types I and II media}

The condition, Tr~$\mathbb\Delta=0$, can be put in explicit form
\cite{Azzam} for the components of the optical matrix, $\mathbf M$, 
\begin{equation}
  \begin{array}{rcl}
    0&=&\rho'_{21}+(\frac{k_x}{\omega}+\rho'_{23})F_1+\mu_{23}F_5,\\
    0&=&\rho_{12}+\epsilon_{13}F_4+\rho_{13}F_8,\\
    0&=&\rho'_{12}+\rho'_{13}F_2+\mu_{13}F_6,\\
    0&=&\rho_{21}+(-\frac{k_x}{\omega}+\rho_{23})F_7+\epsilon_{23}F_3,
  \end{array}
\end{equation}
where
\begin{equation}
  \begin{array}{rcl}
    F_1&=&(\rho'_{31}\rho_{33}-\epsilon_{31}\mu_{33})/G,\\
    F_2&=&((\rho'_{32}-\frac{k_x}{\omega})\rho_{33}-
    \epsilon_{32}\mu_{33})/G,\\  
    F_3&=&(\mu_{31}\rho_{33}-\rho_{31}\mu_{33})/G,\\
    F_4&=&(\mu_{32}\rho_{33}-(\rho_{32}+\frac{k_x}{\omega})\mu_{33})/G,\\
    F_5&=&(\rho'_{33}\epsilon_{31}-\epsilon_{33}\rho'_{31})/G,\\
    F_6&=&(\rho'_{33}\epsilon_{32}-\epsilon_{33}(\rho'_{32}-
    \frac{k_x}{\omega}))/G,\\ 
    F_7&=&(\rho'_{33}\rho_{31}-\epsilon_{33}\mu_{31})/G,\\
    F_8&=&(\rho'_{33}(\rho_{32}+\frac{k_x}{\omega})-
    \epsilon_{33}\mu_{32})/G,\\ 
    G&=&\epsilon_{33}\mu_{33}-\rho'_{33}\rho_{33},
  \end{array}
\end{equation}
therefore, we can discard as candidates for our formalism, media with
non isotropic optical activity, as well as with dielectric or
permeability tensors with non-zero elements in their third column/row
off the diagonal. Despite this restriction, there is still many
systems suitable for this formalism. Here we provide some examples.  

The first example consists of a birefringent uniaxial medium with its
optical axis perpendicular to the axis of propagation and rotated from
the plane of incidence, in this case,
$\mathbf\rho=\mathbf\rho'=\mathbb 0$ and 
$\mathbf\mu=\mathbf\mu_v\mathbf I$, where $\mathbf I$ is the identity
$3\times3$-matrix, and 
the form of the $\mathbb \Delta$ matrix is \cite{Berreman-1972,
  Sosnowski} 
\begin{equation}\label{Gamma}
\mathbb \Delta=\left(\begin{array}{cccc}
0&\mu_v-\frac{k_x^2}{\omega^2\epsilon_o} &0&0\\
 \epsilon_{xx}&0&\epsilon_{xy}&0\\
 0&0&0&\mu_v\\
\epsilon_{yx}&0&\epsilon_{yy}-\frac{k_x^2}{\omega^2\mu_v}&0 
\end{array}\right),
\end{equation}
which is of the type I, where
\begin{equation}
  \begin{array}{cc}
    \epsilon_{xx}&=\epsilon_e\cos^2\psi+\epsilon_o\sin^2\psi,\\
    \epsilon_{yy}&=\epsilon_o\cos^2\psi+\epsilon_e\sin^2\psi,\\
    \epsilon_{xy}&=\epsilon_{yx}=\frac{1}{2}(\epsilon_o-\epsilon_e)
    \sin{2\psi},  
    \end{array}
\end{equation}
$\epsilon_v$ and $\mu_v$ are the vacuum permitivity and permeability,
respectively, $n_o=\sqrt{\epsilon_o/\epsilon_v}$, and
$n_e=\sqrt{\epsilon_e/\epsilon_v}$, are the ordinary and extraordinary
refractive indexes, respectively, and $\psi$ is the angle of rotation
of the optical axis from the $x$-axis.
The corresponding eigenvalues are \cite{Sosnowski,Lekner-1991},
\begin{equation}
  k_{\ell}=\pm\frac{\omega J}{c}, \pm\frac{\omega I}{c}, \quad \ell=1,2,3,4,
\end{equation}
where,
\begin{equation}
  J=\sqrt{n_o^2-n_0^2\sin^2\theta_0},
\end{equation}
and
\begin{equation}
  I=\sqrt{n_e^2-n_0^2\sin^2\theta_0\epsilon_{xx}/\epsilon_o}, 
\end{equation}
with eigenvector matrix $\mathbb V=(\vec V_1, \vec V_2, \vec V_3,\vec
V_4)$ \cite{Sosnowski},
\begin{equation}\label{Psimatrix}
  \mathbb V=
  \left(\begin{array}{cccc}
    \sin\psi&\cos\psi&\sin\psi&\cos\psi\\
    \frac{n_o^2\sin\psi}{JZ_v}&\frac{n_o^2I\cos\psi}{J^2Z_v}&
    -\frac{n_o^2\sin\psi}{JZ_v}&  -\frac{n_o^2I\cos\psi}{J^2Z_v}\\   
    \cos\psi&-\frac{n_o^2\sin\psi}{J^2}&\cos\psi&
    -\frac{n_o^2\sin\psi}{J^2}\\ 
   \frac{J\cos\psi}{Z_v}&  -\frac{n_o^2I\sin\psi}{J^2Z_v}&
   -\frac{J\cos\psi}{Z_v}& \frac{n_o^2I\sin\psi}{J^2Z_v}
\end{array}\right), 
\end{equation}
where $Z_v$ is the vacuum impedance.   
The corresponding inverse matrix, $\mathbb V^{-1}$, is 
\begin{equation}\label{inveigenvectors}
  \mathbb V^{-1}=\frac{J^2Z_v}{2I(n_o^2\sin^2\psi+J^2\cos^2\psi)}
  \left(\begin{array}{cccc}  
    \frac{n_o^2I\sin\psi}{J^2Z_v}&\frac{I\sin\psi}{J}&
    \frac{I\cos\psi}{Z_v}&\frac{I\cos\psi}{J}\\
    \frac{I\cos\psi}{Z_v}&\frac{J^2\cos\psi}{n_o^2}&
    -\frac{I\sin\psi}{Z_v}& -\sin\psi\\
    \frac{n_o^2I\sin\psi}{J^2Z_v}&-\frac{I\sin\psi}{J}&
    \frac{I\cos\psi}{Z_v}&-\frac{I\cos\psi}{J}\\
    \frac{I\cos\psi}{Z_v}&-\frac{J^2\cos\psi}{n_o^2}&
    -\frac{I\sin\psi}{Z_v}&\sin\psi
  \end{array}\right).
\end{equation}
Notice both expressions, are in agreement with Eqs.~(\ref{vmatrix})
and Eqs.~(\ref{umatrix}). Also notice when $\psi=0$,
the off-diagonal $2\times2$ block matrices of the differential
propagation matrix, $\mathbb\Delta$, become null and the eigenvectors
$\vec V_1$ and $\vec V_2$ become orthogonal, therefore the $\mathbb V$
matrix can be written in either form, in this case in the form
corresponding to type I matrices. However, the non-degeneracy
condition of Eq.~(\ref{extracond}) is still fulfilled as long as
$n_o\ne~n_e$.    

With little modifications, we have a second example that consists of an
isotropic medium, that in the presence of an external magnetic field
along the $z$-axis, exhibits the Faraday effect, in this case \cite{Born},     
\begin{equation}\label{Gamma_Faraday}
\mathbb \Delta=\left(\begin{array}{cccc}
0&\mu_v-\frac{k_x^2}{\omega^2\epsilon} &0&0\\
 \epsilon&0&i\gamma&0\\
 0&0&0&\mu_v\\
-i\gamma&0&\epsilon-\frac{k_x^2}{\omega^2\mu_v}&0 
\end{array}\right),
\end{equation}
which has the same structure than the birefringent uniaxial medium and
therefore is of the type I.
The corresponding eigenvalues are 
\begin{equation}
  k_{\ell}=\frac{\omega p_\ell}{c}, \quad \ell=1,2,3,4,
\end{equation}
where
\begin{equation}
  p_\ell=\sqrt{J\left(J\pm\frac{\gamma}{\sqrt{\epsilon\epsilon_v}}\right)}, \quad \ell=1,2, 
\end{equation}
\begin{equation}
  J=\sqrt{n^2-n_0^2\sin^2\theta_0},
\end{equation}
$n=\sqrt{\epsilon/\epsilon_v}$ is the isotropic refraction index, and
$p_{\ell+2}=-p_\ell$. The eigenvector matrix, $\mathbb V=(\vec 
V_1, \vec V_2, \vec V_3,\vec V_4)$, is
\begin{equation}\label{Psimatrix-Faraday}
  \mathbb V=
  \left(\begin{array}{cccc}
    1&1&1&1\\
    \frac{n^2p_1}{J^2Z_v}&\frac{n^2p_2}{J^2Z_v}&
    -\frac{n^2p_1}{J^2Z_v}&-\frac{n^2p_2}{J^2Z_v}\\
    -in/J&in/J&-in/J&in/J\\ 
   -i\frac{np_1}{JZ_v}& i\frac{np_2}{JZ_v}&
   i\frac{np_1}{JZ_v}& -i\frac{np_2}{JZ_v},
\end{array}\right).
\end{equation}
The corresponding inverse matrix, $\mathbb V^{-1}$, is 
\begin{equation}\label{inveigenvectors-Faraday}
  \mathbb V^{-1}=\frac{1}{4}
  \left(\begin{array}{cccc}  
    1& \frac{J^2Z_v}{n^2p_1}&iJ/n&i\frac{JZ_v}{np_1}\\
    1& \frac{J^2Z_v}{n^2p_2}&-iJ/n&-i\frac{JZ_v}{np_2}\\
    1& -\frac{J^2Z_v}{n^2p_1}&iJ/n&-i\frac{JZ_v}{np_1}\\
    1& -\frac{J^2Z_v}{n^2p_2}&-iJ/n&i\frac{JZ_v}{np_2},
  \end{array}\right).
\end{equation}
Notice again both expressions, are in agreement with
Eqs.~(\ref{vmatrix}) and Eqs.~(\ref{umatrix}), which
  in this particular case guarantees the breaking of the time reversal
  symmetry \cite{Haldane-2008}.
 
A third example consists of an anisotropic magnetic medium, with
permeability tensor
\begin{equation}\label{magnetic_tensor}
\mathbf \mu=\left(\begin{array}{ccc}
\mu_{11}&\mu_{12}&0\\
\mu_{21}&\mu_{22}&0\\
 0&0&\mu_{33}
\end{array}\right),
\end{equation}
$\mathbf\epsilon=\tilde\epsilon\mathbf I$, where $\tilde \epsilon$ is a
scalar and $\mathbf\rho=\mathbf\rho'=\mathbb 0$,
in this case,
\begin{equation}\label{Gamma_magnetic}
\mathbb \Delta=\left(\begin{array}{cccc}
0&\mu_{22}-\frac{k_x^2}{\omega^2\tilde\epsilon} &0&-\mu_{21}\\
 \tilde\epsilon&0&0&0\\
 0&-\mu_{12}&0&\mu_{11}\\
0&0&\tilde\epsilon-\frac{k_x^2}{\omega^2\mu_{33}}&0 
\end{array}\right),
\end{equation}
which is also of the type I. We omit its eigenvalues and eigenvectors.

Finally, our last example consists of a medium with isotropic
dielectric, permeability and optical tensors,
i.e. $\mathbf\epsilon=\tilde\epsilon\mathbf I$, where $\tilde
\epsilon$ is a scalar, $\mathbf\mu=\mu_v\mathbf I$,
$\mathbf\rho'=-i\tilde\alpha\mathbf I$, and
$\mathbf\rho=i\tilde\alpha\mathbf I$, where $\tilde\alpha$ is a
scalar
\cite{Treyakov-1998,Oldano-1999,Wade-2020}. In this
case,  
\begin{equation}\label{Gamma_iso}
\mathbb \Delta=\left(\begin{array}{cccc}
0&\mu_v\left(1-\frac{k_x^2}{\omega^2(\tilde\epsilon\mu_v-\tilde\alpha^2)}\right)
&-i\tilde\alpha\left(1+\frac{k_x^2}{\omega^2(\tilde\epsilon\mu_v-\tilde\alpha^2)}\right)&
0\\ 
 \tilde\epsilon&0&0&-i\tilde\alpha\\
 i\tilde\alpha&0&0&\mu_v\\
0&i\tilde\alpha\left(1+\frac{k_x^2}{\omega^2(\tilde\epsilon\mu_v-\tilde\alpha^2)}\right)&\tilde\epsilon\left(1-\frac{k_x^2}{\omega^2(\tilde\epsilon\mu_v-\tilde\rho\tilde\rho')}\right)
&0 
\end{array}\right),
\end{equation}
which is of the type II. The corresponding  eigenvalues are 
\begin{equation}
  k_{\ell}=\frac{\omega p_\ell}{c}, \quad \ell=1,2,3,4,
\end{equation}
where
\begin{equation}
  p_\ell=\sqrt{\left(\tilde\alpha c\pm n\right)^2-n_0^2\sin^2\theta_0},
  \quad \ell=1,2,  
\end{equation}
$n=\sqrt{\epsilon/\epsilon_v}$ is the isotropic refraction index,
and $p_{\ell+2}=-p_\ell$, with eigenvector matrix, $\mathbb V=(\vec
V_1, \vec V_2, \vec V_3,\vec V_4)$, 
\begin{equation}\label{op-act}
  \mathbb V=
  \left(\begin{array}{cccc}
    1&1&1&1\\
    \frac{n(n+\tilde\alpha c)}{Z_vp_1}&\frac{n(n-\tilde\alpha c)}{Z_vp_2}&
  -\frac{n(n+\tilde\alpha c)}{Z_vp_1}&-\frac{n(n-\tilde\alpha
    c)}{Z_vp_2}\\
  i\frac{(n+\tilde\alpha c)}{p_1}&i\frac{(\tilde\alpha c-n)}{p_2}&
  -i\frac{(n+\tilde\alpha c)}{p_1}&-i\frac{(\tilde\alpha c-n)}{p_2}\\
    in/Z_v&-in/Z_v&in/Z_v&-in/Z_v
\end{array}\right). 
\end{equation}
The corresponding inverse matrix, $\mathbb V^{-1}$, is 
\begin{equation}\label{inveigenvects-optact}
  \mathbb V^{-1}=\frac{1}{4}
  \left(\begin{array}{cccc}  
    1& \frac{Z_vp_1}{n(n+\tilde\alpha c)}&-i\frac{p_1}{n+\tilde\alpha
      c}&-iZ_v/n\\
    1& \frac{Z_vp_2}{n(n-\tilde\alpha c)}&-i\frac{p_2}{\tilde\alpha
       c-n}&iZ_v/n\\
    1& -\frac{Z_vp_1}{n(n+\tilde\alpha c)}&i\frac{p_1}{n+\tilde\alpha
      c}&-iZ_v/n\\
    1& -\frac{Z_vp_2}{n(n-\tilde\alpha c)}&i\frac{p_2}{\tilde\alpha
       c-n}&iZ_v/n
  \end{array}\right).
\end{equation}
Notice once again, that both expressions, are in agreement with
Eqs.~(\ref{vmatrix}) and Eqs.~(\ref{umatrix}), which in this
particular case and unlike the Faraday rotator, guarantees the time 
reversal symmetry \cite{Berreman-1972,Drude-1965,Haldane-2008}.

\end{document}